\documentclass[Physsubmission, Phys]{SciPost}
\hypersetup{
    colorlinks,
    linkcolor={red!50!black},
    citecolor={blue!50!black},
    urlcolor={blue!80!black}
}

\usepackage[bitstream-charter]{mathdesign}
\usepackage{siunitx}
\DeclareSymbolFont{usualmathcal}{OMS}{cmsy}{m}{n}
\DeclareSymbolFontAlphabet{\mathcal}{usualmathcal}

 \newcommand{\dsk}{DS\textendash20k} 

\begin{document}

%\linenumbers

\begin{center}{\Large \textbf{
The DarkSide\textendash 20k TPC and Underground Argon Cryogenic System\\
}}\end{center}

% TODO: write the author list here. Use initials + surname format.
% Separate subsequent authors by a comma, omit comma at the end of the list.
% Mark the corresponding author with a superscript *.
\begin{center}
T.~N.~Thorpe\textsuperscript{1} on behalf of the DS\textendash 20k Collaboration,
\end{center}

\begin{center}
{\bf 1} Department of Physics and Astronomy, University of California \textendash \ Los Angeles, 475 Portola Plaza, Los Angeles, CA 90095
\\

tnthorpe@g.ucla.edu
\end{center}

\begin{center}
\today
\end{center}

% For convenience during refereeing (optional),
% you can turn on line numbers by uncommenting the next line:
%\linenumbers

\definecolor{palegray}{gray}{0.95}
\begin{center}
\colorbox{palegray}{
  \begin{tabular}{rr}
  \begin{minipage}{0.1\textwidth}
    \includegraphics[width=30mm]{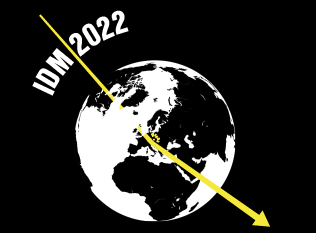}
  \end{minipage}
  &
  \begin{minipage}{0.85\textwidth}
    \begin{center}
    {\it 14th International Conference on Identification of Dark Matter}\\
    {\it Vienna, Austria, 18-22 July 2022} \\
    \doi{10.21468/SciPostPhysProc.?}\\
    \end{center}
  \end{minipage}
\end{tabular}
}
\end{center}

\section*{Abstract}
{\bf
% TODO: write your abstract here.
DarkSide\textendash 20k (\dsk) will exploit the physical and chemical properties of liquid argon (LAr) housed within a large dual-phase time project chamber (TPC) in its direct search for dark matter.  The TPC will utilize a compact, integrated design with many novel features to enable the \SI{20}{t} fiducial volume of underground argon.  Underground Argon (UAr) is sourced from underground CO$_2$ wells and depleted in the radioactive isotope $^{39}$Ar, greatly enhancing the experimental sensitivity to dark matter interactions.  Sourcing and transporting the $\mathcal{O}$(\SI{100}{t}) of UAr for \dsk\ is costly, and a dedicated single-closed-loop cryogenic system has been designed, constructed, and tested to handle the valuable UAr.  We present an overview of the \dsk\ TPC design and the first results from the UAr cryogenic system.   
}

\vspace{10pt}
\noindent\rule{\textwidth}{1pt}
\tableofcontents\thispagestyle{fancy}
\noindent\rule{\textwidth}{1pt}
\vspace{10pt}

\section{Introduction}
\label{sec:intro}
% TODO: write your article here.
DarkSide\textendash 20k (\dsk ) is an upcoming direct dark matter search experiment and will be located in the underground laboratory of the Gran Sasso National Laboratory (LNGS) in Assergi, IT.  \dsk\ is designed to observe dark matter particles scattering from argon atoms in the liquid argon (LAr) target. The measureable signal from Weakly Interacting Massive Particle (WIMP) dark matter scattering is a nuclear recoil, depositing energies up to $\mathcal{O}$(100) \SI{}{keV} within the LAr.  The detector is designed to operate for a minimum of \SI{10}{years}  while maintaining a negligible instrumental background level in the WIMP search region of interest.    

The core of the \dsk\ experiment is a dual\textendash phase Time Projection Chamber (TPC) which will be filled and surrounded with low\textendash radioactivity Underground Argon (UAr).  The level of the $\beta$\textendash emitting radioactive isotope, $^{39}$Ar in UAr is less ($\approx 1400 \times$) than that of standard argon of atmospheric origin (AAr), as demonstrated \cite{Agnes:2018ep} by the predecessor experiment DarkSide\textendash 50.  The fiducial volume of \dsk\ is \SI{20}{t}, which gives a minimum design exposure of \SI{200}{t \times years} using UAr.  We briefly overview the \dsk\ TPC and UAr cryogenic system design, and discuss the first cryogenic testing results. 

\section{Overview of the DarkSide\textendash 20k Time Projection Chamber (TPC)}
\label{sec:tpc}

\begin{figure}[h]
\centering
\includegraphics[width=0.6\textwidth]{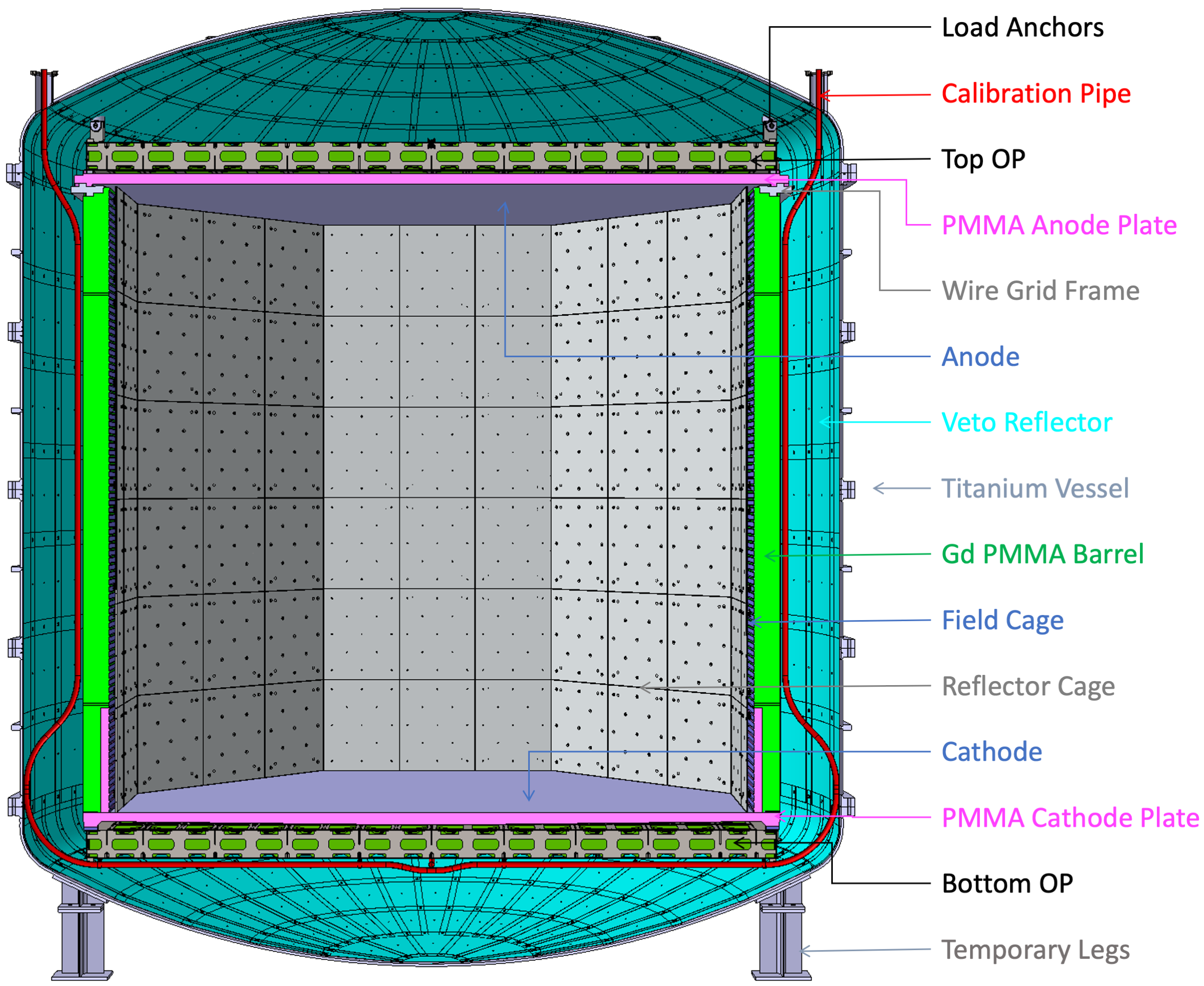}
\includegraphics[width=0.35\textwidth]{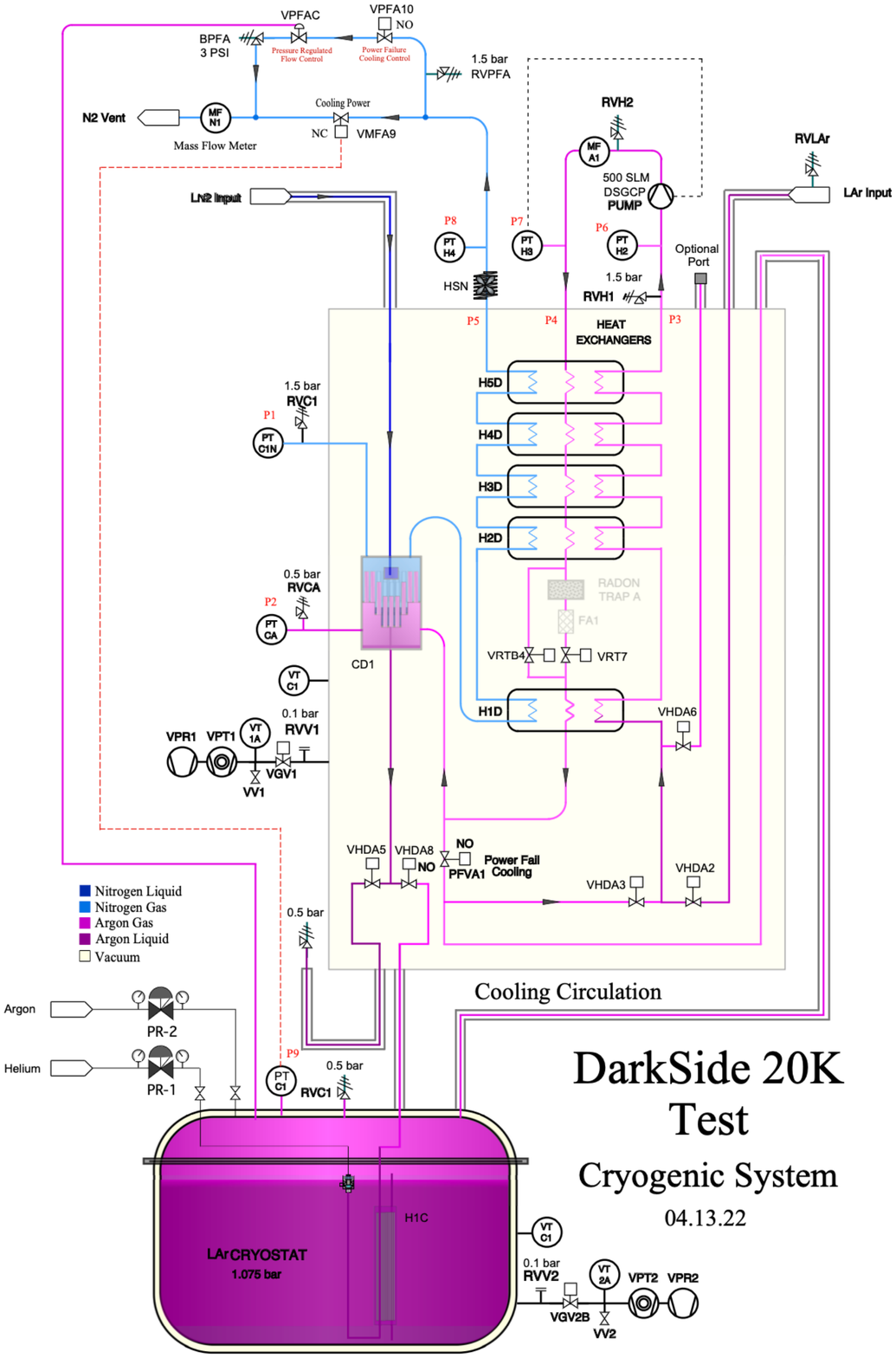}
\caption{Left: Cross\textendash section view of the \dsk \ TPC and neutron veto within the stainless steel vessel which will house the Underground Argon (UAr).  Right: Piping and Instrumentation Diagram (P\&ID) of the \dsk\ Underground argon (UAr) cryogenic testbed at CERN.  Blue (pink) depicts the nitrogen (argon) pathways.}
\label{fig:tpc_cryo}
\end{figure}

The \dsk\ TPC and active neutron veto are two sides of the same mechanical object, which is pictured in the left panel of Fig. \ref{fig:tpc_cryo}.  Eight gadolinium (Gd) loaded PMMA (acrylic) panels (colored in green) will be mounted together to form the \emph{TPC barrel} with the Gd serving as a high cross section target for thermal neutrons.  The panels measure $\SI{365}{cm} \times \SI{163}{cm} \; (l \times w)$ and are \SI{15}{cm} thick except for the bottom section that the cathode barrel fits into, where they measure \SI{10}{cm} thick.  The \emph{anode} and \emph{cathode} plates are pure PMMA pieces coated with Clevios$^{TM}$ and 1,1,4,4-tetraphenyl-1,3-butadiene (TPB).  Clevios$^{TM}$ is a commercial conductive polymer and will be used to define most of the electrical potentials in the TPC.  The TPB coating has a nominal application density of $\SI{200}{\mu g / cm^2}$, and is used to wavelength shift the \SI{128}{nm} argon scintillation light to $\approx$ \SI{420}{nm} to be efficiently detected by the Silicon Photomultiplier (SiPM)\textendash based readout.  Mounted on the interior panel walls is the \emph{reflector cage} composed of \SI{4}{mm} thick PMMA panels combined with TPB\textendash coated \SI{50}{\mu m} thick Enhanced Specular Reflector (ESR) foils. 

%The anode plate is \SI{5}{cm} thick, has a small \SI{13}{mm} ``lip" to contain the gas pocket, and eight holes machined around the edge connecting to the UAr cryogenic system to control the flow into and out of the gas pocket region.  The cathode plate has a sloping exterior surface, hence a varying thickness of \SI{7}{cm} at the center to \SI{5}{cm} at the edges, to avoid trapping any gas bubbles beneath.  The cathode plate has a bonded barrel section which fits inside of the TPC barrel such that the inner surfaces containing the machined Clevios$^{TM}$ coated grooves (field cage rings) align to form the electric field (drift) cage.  The field cage connections are made using redundant \emph{resistor links}, which are Kapton PCBs with copper traces and resistors imbedded.  The cathode is nominally held a voltage of \SI{-73.38}{kV} while the anode is grounded.  The \emph{wire grid} consists of an octagonal stainless steel frame sitting on top of the TPC barrel (below the anode plate) inset with $\SI{150}{\mu m}$ wires spaced \SI{3}{mm} apart held at a nominal voltage of \SI{-3.78}{kV} to create the extraction field.  Since the wire grid frame and acrylic components have different thermal contraction coefficients, \emph{concentricity guides} will be machined into every surface where differential thermal contraction occurs, ensuring that the detector as a whole contracts towards the center without obstruction.

The SiPM\textendash based readouts take the form of large octagonal planes positioned above (below) the cathode (anode) plate.  The readout is divided into $20 \times \SI{20}{cm^2}$ electromechanical \emph{Photo Detector Units (PDUs)}, each with four readout channels, which are mounted onto stainless steel frames forming the \emph{Optical Planes (OPs)}.  Behind the readout layers, held within the frames, \SI{15}{cm} thick Gd\textendash loaded PMMA pieces form the endcaps of the active neutron veto.  Mounted on the opposing side of the frames, as on the exterior of the Gd\textendash loaded barrel wall panels, are the $20 \times \SI{20}{cm^2}$ \emph{Veto Photo Detector Units (VPDUs)} looking towards the reflective inner surface of the stainless steel vessel (covered with poly(ethylene naphthalate) (PEN) and ESR foils).

\section{The DarkSide\textendash 20k Underground Argon (UAr) Cryogenic System}
\label{sec:cryo}

The \dsk\ UAr cryogenic system builds on the successful design of the system used for DarkSide\textendash 50 \cite{Agnes:2015gu}, which operated for over 8 years.  The UAr system is responsible for maintaining the $\mathcal{O}$(100)\,t of target material for the direct dark matter search, including continuous purification of the gaseous argon and reliquefication.  Here we focus on a scaled down version (right panel of Fig. \ref{fig:tpc_cryo}) that was built and tested within the Cryolab at CERN without the purification loop and radon trap.  The complete system Piping and Instrumentation Diagram (P\&ID) for the \dsk\ experiment is included in Appendix \ref{sec:complete_pid} (Fig. \ref{fig:complete_pid}).  The UAr system design is novel and this \textit{testbed} cryogenic system, which contains the core system to be used in \dsk, was constructed to establish fundamental design function and operational parameters.

\subsection{Design Overview}
\label{cryo_design}

%\begin{figure}[h]
%\centering
%\includegraphics[width=0.5\textwidth]{DS20K II P&ID 6.21.22_crop.pdf}
%\caption{Piping and Instrumentation Diagram (P\&ID) of the \dsk\ Underground argon (UAr) cryogenic testbed at CERN.  Blue (pink) depicts the nitrogen (argon) pathways.}
%\label{fig:pid}
%\end{figure}

The right panel of Fig. \ref{fig:tpc_cryo} details the testbed system used for functionality and performance measurements.  The \emph{coldbox} is a vacuum insulated vessel depicted by the yellowish inset region, and blue (pink) depicts the nitrogen (argon) pathways.  The core of the UAr cryogenic system is the \emph{condenser}, which is the small rectangle labeled as CD1 in the right panel of Fig. \ref{fig:tpc_cryo}.  The condenser consists of 1/2\textendash inch stainless steel tubes capped on one end with the other ends welded to a base\textendash plate with 1/2\textendash inch holes.  Covers are then welded onto both sides of the base\textendash plate creating a two sided object separated by a tubular geometry allowing for heat exchange.  

The condenser principle of operation follows:  The tube openings are faced downwards and gaseous argon (GAr) is flowed to fill the tube volumes.  Liquid nitrogen (LN$_2$) flows steadily onto the top side of the tubes, and latent heat exchange occurs.  The condensed argon moves down the inner surfaces of the tubes and is routed into the cryostat, while the gaseous nitrogen is routed out of the top of the condenser.  Temperatures and pressures are measured throughout the system, and gaseous argon and nitrogen flows are measured at room temperature.

A sophisticated network of parallel plate heat exchangers is employed within the coldbox where gaseous phase heat exchange continues over the entire temperature range from \SI{87}{K} to room temperature.  The exhausted N$_2$ gas can be seen routed on the left side of the heat exchangers.  The argon pathway is a single closed loop which includes the cryostat and, as argon boils, the cold gas is routed to the bottom\textendash right side of the heat exchangers.  A gaseous pump circulates the argon, with the warm gas from the pump being routed down the middle of the heat exchange network.  This routing allows the gaseous argon to be precooled to near liquid temperature before entering the condenser.  This design realizes an extremely efficient system with abundant cooling power being recovered from the intrinsic boiling of LAr in the cryostat. 

\subsection{Preliminary Commissioning Results}
\label{cryo_results}

We present the cooling power recovery efficiency results and describe the detector circulation concept with some first results.

\subsubsection{Cooling Power Recovery Efficiency} 

\begin{figure}[h]
\centering
\includegraphics[width=0.49\textwidth]{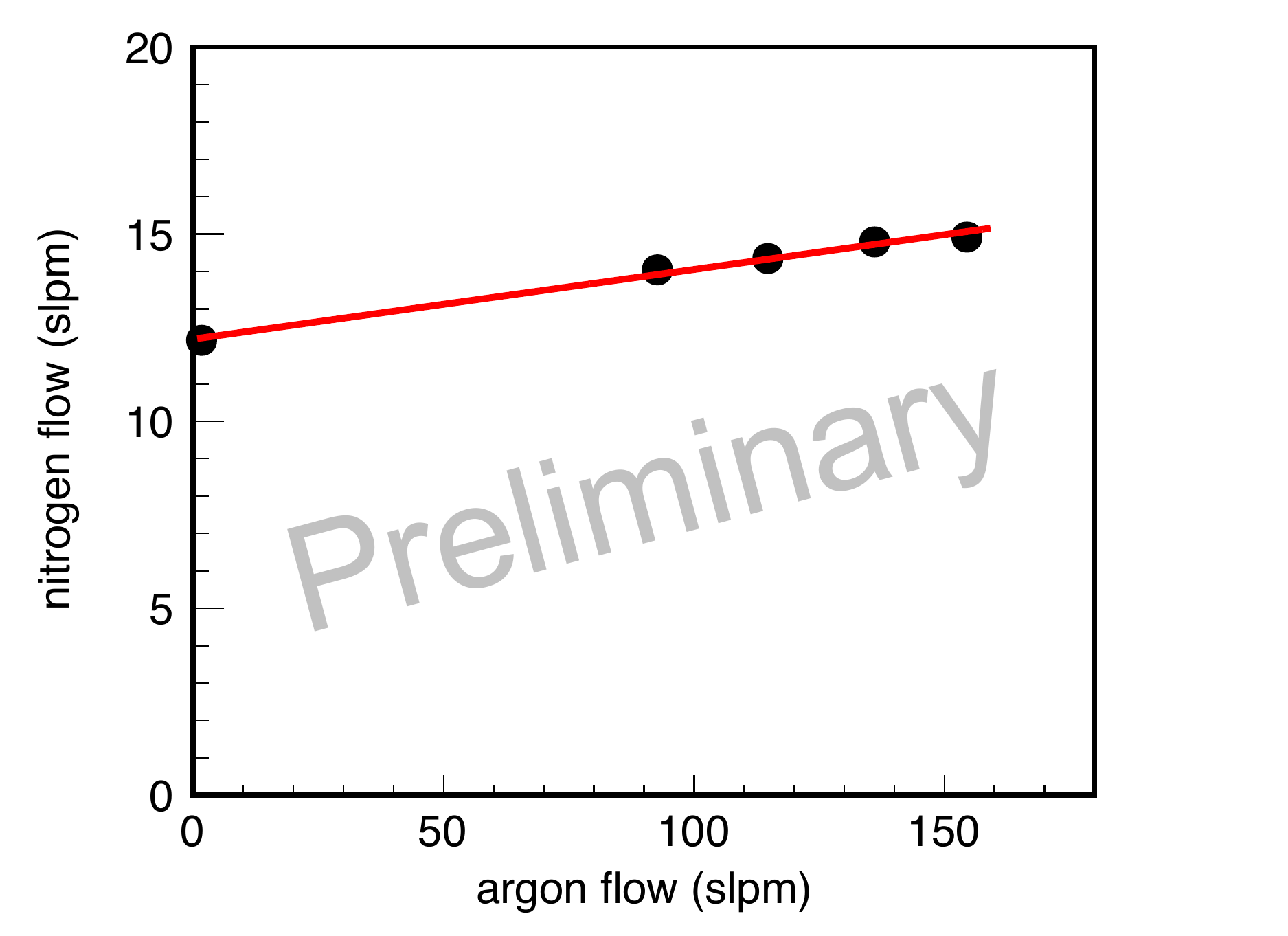}
\includegraphics[width=0.49\textwidth]{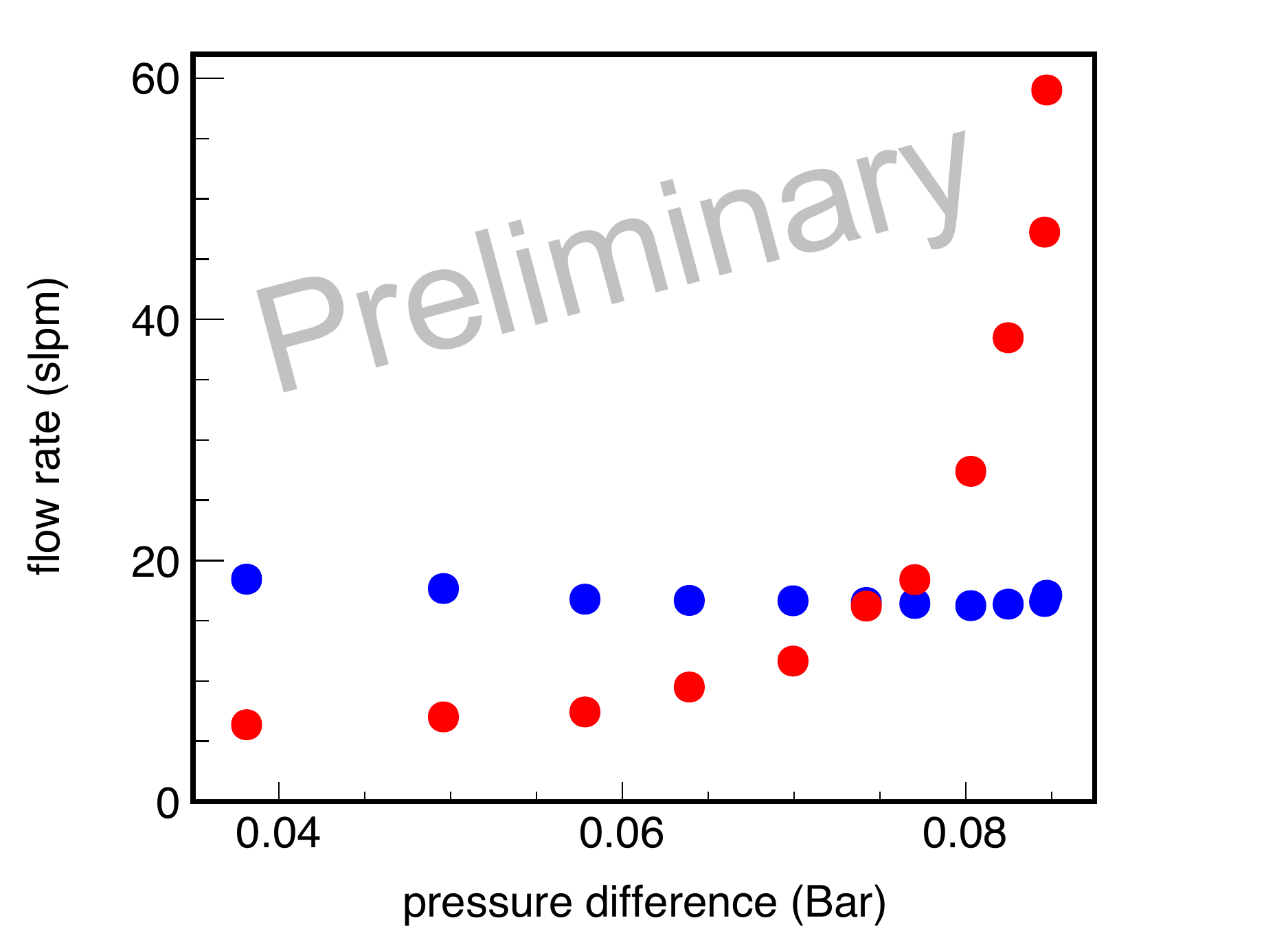}
\caption{Left: Nitrogen mass flow (consumption) versus argon mass flow.  The data points (black) are taken at different argon circulation speeds.  The solid line (red) is a linear $\chi^2$ minimization.  The returned slope is taken as the consumption, which is converted into a cooling power recovery efficiency of at least 99.1\%.  Right: Argon (red data points) and nitrogen (blue data points) flow versus the pressure difference across the heat exchangers.  As the pressure difference increases, the argon flow rises sharply indicating that the gaseous/liquid mixture on one side of the heat exchangers is being condensed.  The heat from this process boils LAr on the opposing side of the heat exchangers, producing gas flow for the detector circulation.  The nitrogen consumption is essentially flat, indicating the process is highly efficient. }
\label{fig:cryo_results}
\end{figure}

Visible inside of the cryostat in the right panel of Fig. \ref{fig:tpc_cryo} (labeled as H1C) is another parallel plate heat exchange system for detector circulation, which is actually two heat exchangers mounted in parallel.  The condenser outlet is routed into one side of heat exchangers, which then passes through a helium\textendash controlled valve into ullage volume of the cryostat.  The helium\textendash controlled valve can be used to create a back pressure on the argon outlet, simulating a head pressure which exists in the \dsk\ experiment.

For this test the helium valve is not used (no pressure besides that in the ullage is set), and gaseous argon is circulated.  The argon circulation pump is set to various speeds for 12\textendash hour increments, allowing the system to become well equilibrated.  In analysis, time periods are chosen where the flows are most stable and linear $\chi^2$ minimizations are performed on the flow data.  The returned fit parameters are used to evaluate the average nitrogen and argon flows over long time periods.  The results (black data points) are shown in the left panel of Fig. \ref{fig:cryo_results} as nitrogen flow (consumption) versus argon flow.  A linear $\chi^2$ minimization is performed on the consumption data, and the returned slope value is regarded as a ``consumption rate", i.e. the amount of nitrogen needed to cool and condense a given amount of argon.

To obtain a value for the cooling power recovery efficiency, we define an efficiency scale in the parameter space of Fig. \ref{fig:cryo_results}.  A zero slope indicates a 100\% efficient system.  A system with 0\% efficiency is defined as the amount of nitrogen (latent heat only) needed to cool and condense argon from room temperature.  While gas is mainly being circulated here, we assume the argon phase change within H1C would be 100\% efficient as it occurs within the LAr bulk (with the heat from this process generating the detector circulation).  Under these assumptions, we determine our cooling power recovery efficiency to be at least 99.1\%.

\subsubsection{Detector Circulation}

The detector circulation is a crucial parameter for gaseous/liquid detectors because the target volume must be purified at a certain rate in order to perform rare\textendash event physics searches.  At a rate of \SI{1000}{slpm}, the $\approx$ \SI{100}{t} of UAr in \dsk\ is replaced roughly every 40 days.  As mentioned in the previous section, a helium\textendash controlled valve is used to simulate the depth that the detector circulation heat exchange system is located within \dsk.  LAr, which has filled the opposing side of the heat exchangers, is boiled to produce the detector circulation for this test.

The helium\textendash controlled valve is pressurized and the circulation pump speed is increased until the argon pressure within the heat exchanger overcomes the helium valve pressure.  Once argon gas is visibly flowing from the outlet (via a camera mounted inside), liquid is not being condensed, and the test is complete.  As the circulation speed is incrementally increased, a pressure difference over the heat exchangers is established and the gaseous/liquid argon mixture coming from the condenser is compressed into LAr.  As mentioned, the heat from this process boils the LAr on the opposing side of the heat exchangers resulting in a gas supply for the detector circulation.  The results are shown in the right panel of Fig. \ref{fig:cryo_results}. 

%\begin{figure}[h]
%\centering
%\includegraphics[width=0.7\textwidth]{cryo_circulation_results.pdf}
%\caption{Argon (red circles) and nitrogen (blue circles) flow versus the pressure across the heat exchangers.  As the pressure difference increases, the argon flow rises sharply indicating that the gaseous/liquid mixture on one side of the heat exchanger is being condensed.  The heat from this process boils LAr on the opposing side of the heat exchangers, producing gas flow for the detector circulation.  The nitrogen consumption is essentially flat, indicating this process is highly effcient. }
%\label{fig:circulation}
%\end{figure}

\section{Conclusion}

The \dsk\ experiment will search for dark matter directly using a dual\textendash phase Time Projection Chamber (TPC), where the target material is argon sourced from underground to exploit the low levels of the radioactive isotope, $^{39}$Ar.  The experiment has entered the construction phase at the Gran Sasso Laboratory (LNGS) in central Italy.  The novel TPC design uses Gd\textendash loaded PMMA (acrylic) panels for the TPC body, mechanical support for the TPC/veto assembly, and target material for the active neutron veto.  The design is in an advanced stage and assembly procedures are being finalized.  The core of the Underground Argon (UAr) cryogenic system has been successfully tested at CERN, and the system is now being shipped to LNGS.  A cooling power recovery efficiency of 99.1\% was measured, and a highly efficient detector circulation method was demonstrated.   

\section*{Acknowledgements}

The DarkSide Collaboration would like to thank LNGS, LSC, SNOLAB, Boulby and their staff for invaluable technical and logistical support. We acknowledge the financial support by the Spanish Ministry of Science and Innovation PID2019-109374GB-I00 (MICINN), “Unidad de Excelencia María de Maeztu: CIEMAT - Física de partículas”
(Grant MDM2015- 0509), the U. S. National Science Foundation (NSF) (Grants No. PHY- 0919363, No. PHY1004054, No. PHY-1004072, No. PHY-1242585, No. PHY-1314483, No. PHY- 1314507, associated collaborative
grants, No. PHY-1211308, No. PHY-1314501, No. PHY-1455351 and No. PHY-1606912, as well as Major Research
Instrumentation Grant No. MRI-1429544), the Italian Istituto Nazionale di Fisica Nucleare (Grants from Italian
Minis- tero dell’Istruzione, Università, e Ricerca Progetto Premiale 2013 and Commissione Scientific Nazionale II),
the Natural Sciences and Engineering Research Council of Canada, SNOLAB, and the Arthur B. McDonald Canadian Astroparticle Physics Research Institute, LabEx UnivEarthS (ANR-10-LABX-0023 and ANR-18- IDEX-0001),
the São Paulo Research Foundation (Grant FAPESP-2017/26238-4), and the Russian Science Foundation Grant No.
16-12-10369, the Polish National Science Centre (Grant No. UMO-2019/33/B/ST2/02884), the Foundation for Polish Science (Grant No. TEAM/2016-2/17), the International Research Agenda Programme AstroCeNT (Grant No.
MAB/2018/7) funded by the Foundation for Polish Science from the European Regional Development Fund, the Science and Technology Facilities Council, part of the United Kingdom Research and Innovation, and The Royal Society (United Kingdom). I.F.M.A is supported in part by Conselho Nacional de Desen- volvimento Científico e Tecnológico
(CNPq). We also wish to acknowledge the support from Pacific Northwest National Laboratory, which is operated by
Battelle for the U.S. Depart- ment of Energy under Contract No. DE-AC05-76RL01830.

% TODO: include author contributions
%\paragraph{Author contributions}
%This is optional. If desired, contributions should be succinctly described in a single short paragraph, using author initials.

% TODO: include funding information
%\paragraph{Funding information}
%Authors are required to provide funding information, including relevant agencies and grant numbers with linked author's initials. Correctly-provided data will be linked to funders listed in the \href{https://www.crossref.org/services/funder-registry/}{\sf Fundref registry}.

\begin{appendix}

\section{\dsk\ Underground Argon (UAr) Cryogenic System P\&ID}
\label{sec:complete_pid}

\begin{figure}[h!]
\centering
\includegraphics[width=\textwidth]{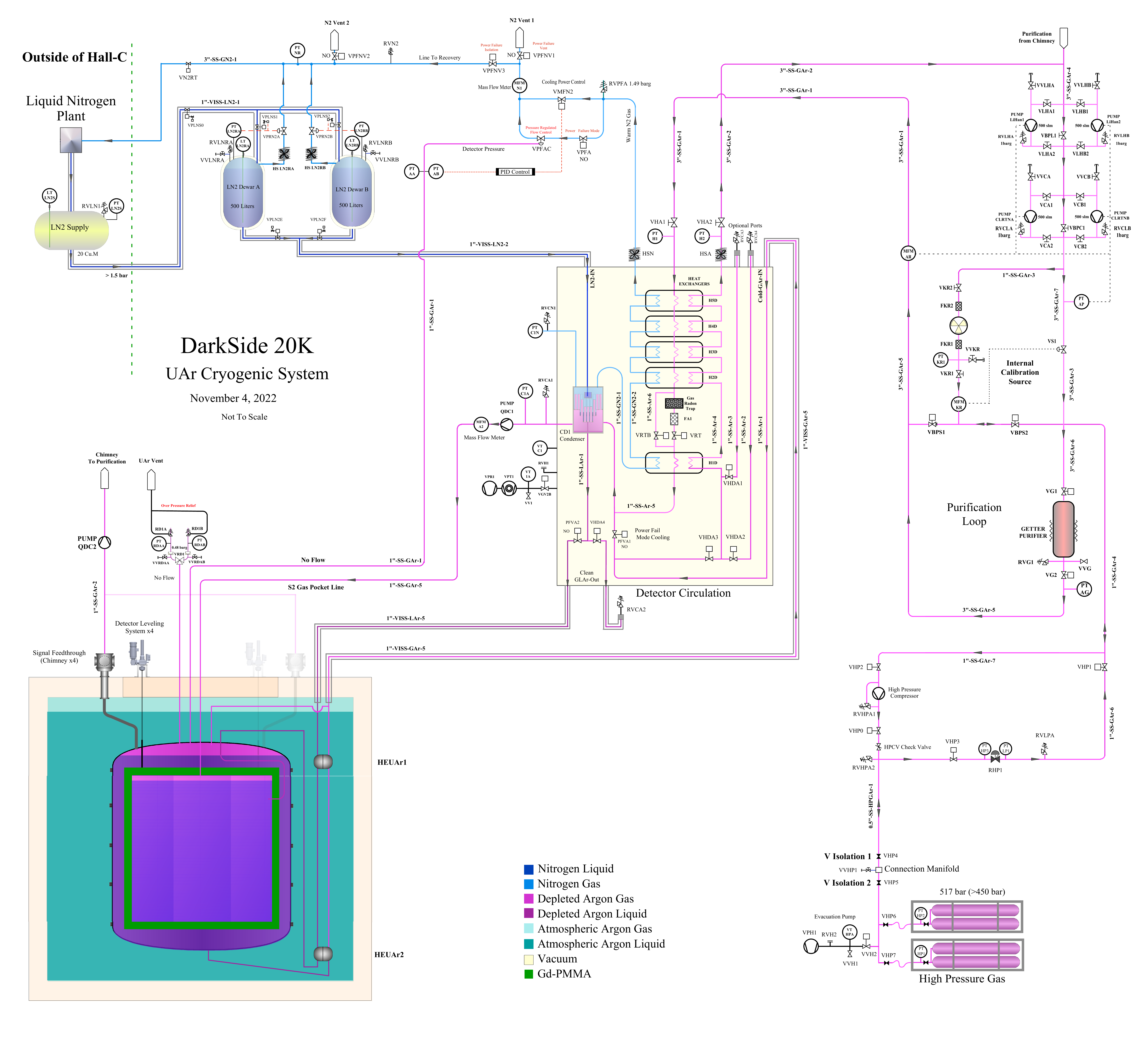}
\caption{Piping and Instrumentation Diagram (P\&ID) of the Underground argon (UAr) cryogenic system for the \dsk\ experiment.  The liquid nitrogen and UAr supply systems, along with the purification loop are included.}
\label{fig:complete_pid}
\end{figure}

%\section{About references}
%Your references should start with the comma-separated author list (initials + last name), the publication title in italics, the journal reference with volume in bold, start page number, publication year in parenthesis, completed by the DOI link (linking must be implemented before publication). If using BiBTeX, please use the style files provided  on \url{https://scipost.org/submissions/author_guidelines}. If you are using our \LaTeX template, simply add
%\begin{verbatim}
\bibliography{ds.bib}
%\end{verbatim}
%at the end of your document. If you are not using our \LaTeX template, please still use our bibstyle as
%\begin{verbatim}
\bibliographystyle{SciPost_bibstyle}
%\end{verbatim}
%in order to simplify the production of your paper.
\end{appendix}

% TODO:
% Provide your bibliography here. You have two options:

% FIRST OPTION - write your entries here directly, following the example below, including Author(s), Title, Journal Ref. with year in parentheses at the end, followed by the DOI number.
%\begin{thebibliography}{99}
%\bibitem{1931_Bethe_ZP_71} H. A. Bethe, {\it Zur Theorie der Metalle. i. Eigenwerte und Eigenfunktionen der linearen Atomkette}, Zeit. f{\"u}r Phys. {\bf 71}, 205 (1931), \doi{10.1007\%2FBF01341708}.
%\bibitem{arXiv:1108.2700} P. Ginsparg, {\it It was twenty years ago today... }, \url{http://arxiv.org/abs/1108.2700}.
%\end{thebibliography}

%\bibliography{ds.bib}
 %\bibliographystyle{SciPost_bibstyle} % Include this style file here only if you are not using our template
%\bibliography{SciPost_Example_BiBTeX_File.bib}

\nolinenumbers

\end{document}